\documentclass[preprint,showpacs,nofootinbib,aps,amsmath,amssymb,tightenlines]{revtex4}

\usepackage{amsmath,amssymb,amsbsy,latexsym,graphics}

\def\R{{\mathbb R}}

\def\man{{\cal M}}

\def\IH{{\Delta}}
\def\d{{\rm d}}
\def\f{\frac}
\def\H{{\cal H}}
\def\lp{\ell_{\rm Pl}}
\def\g{\gamma}
\def\S{\Sigma}
\def\SU(2){{\rm SU(2)}}
\def\U(1){{\rm U(1)}}
\def\Hk{{\H_{\rm kin}}}
\def\Hp{{\H_{\rm phy}}}
\def\k{{(k)}}

\def\ba{\begin{eqnarray}}
\def\ea{\end{eqnarray}}
\def\be{\begin{equation}}
\def\ee{\end{equation}}

\newcommand{\nt}[1]{\mathring{#1}} 

\preprint{\vbox{\baselineskip=12pt
 \rightline{gr-qc/0412nnn}
 \rightline{IGPG 2004-11/9} }}

\begin{document}


\title{Quantum horizons and black hole entropy:\\
Inclusion of distortion and rotation}

\author{Abhay\ Ashtekar}
\email{ashtekar@gravity.psu.edu}
\author{Jonathan\ Engle}
\email{engle@gravity.psu.edu}
\author{Chris\ Van Den Broeck}
\email{vdbroeck@gravity.psu.edu}

\affiliation{Institute for Gravitational Physics and Geometry,\\
    Physics Department, Penn State, University Park, PA 16802,
    USA}


\begin{abstract}

Equilibrium states of black holes can be modelled by isolated
horizons. If the intrinsic geometry is spherical, they are called
type I while if it is axi-symmetric, they are called type II. The
detailed theory of geometry of \emph{quantum} type I horizons and
the calculation of their entropy can be generalized to type II,
thereby including arbitrary distortions and rotations. The leading
term in entropy of large horizons is again given by $1/4$th of the
horizon area for the \emph{same} value of the Barbero-Immirzi
parameter as in the type I case. Ideas and constructions
underlying this extension are summarized.

\end{abstract}

 \maketitle

\emph{Basic ideas}: Isolated horizons (IHs) provide a quasi-local
framework to describe black holes which are themselves in
equilibrium but in space-times whose exterior regions may carry
time-dependent fields and geometry \cite{akrev}. The zeroth and
the first laws of black hole mechanics of classical general
relativity were first established for globally stationary black
holes \cite{bch}. However, they extend to all IHs \cite{afk,abl2}.
The thermodynamic entropy is again given by $1/4$th the horizon
area (provided matter is minimally coupled to gravity). These
results hold not just for the Kerr-Newman family but also for
astrophysically realistic black holes which may be distorted. It
is natural to ask if a quantum gravity description of the IH
geometry can lead to a statistical mechanical calculation of
entropy of this diverse family of black holes.

In the globally stationary situation, black holes without external
influences are completely characterized by their mass (or horizon
area%
\footnote{In the classical isolated horizon framework, area,
angular momentum and charges turn out to be primary; mass is a
\emph{derived} quantity, defined in terms of them
\cite{afk,abl2}.}),
spin and possible charges associated with gauge fields and
dilatons which, however, will be ignored in most of this brief
presentation. The entropy of a black hole with \emph{fixed} mass
and spin is given by $1/4$ the horizon area. In the quasi-local
context of IHs, mass and spin do not suffice to characterize a
time independent horizon geometry. One needs an infinite set of
multipoles \cite{aepv} to capture the distortions in the mass and
angular momentum distribution on the horizon induced, e.g., by
external matter rings, which are ignored by fiat in the black hole
uniqueness theorems.

We will begin by briefly recalling the notion of IHs and their
multipoles. Then, we will sketch the essential features of a
Hamiltonian framework for the sector of general relativity
consisting of space-times which admit an isolated horizon with
fixed multipoles, carry out a non-perturbative quantization using
ideas from quantum geometry \cite{alrev,crbook,ttbook} and finally
calculate the number of microstates of the resulting quantum
horizon. This strategy is the same as that used in
\cite{abck,ack,abk,dl,km} for the simplest (type I) isolated
horizons whose \emph{intrinsic} geometry is spherical.\\

\emph{Isolated horizons}: The precise notion of an isolated
horizon (IH) is arrived at by extracting from the definition of a
Killing horizon the minimum conditions necessary for black hole
thermodynamics. More precisely, one begins with a null,
3-dimensional sub-manifold $\IH$, topologically $S^2\times \R$ and
with null normal $\ell$, of a 4-dimensional space-time $(\man,
g)$, . Its \emph{intrinsic} geometry is coded in a pair $(q, V)$
consisting of a `metric' $q$ of signature 0,+,+ and a
complex-valued $U(1)$ connection $V$ on the spin-bundle of any of
its 2-sphere cross-sections $S$. $\IH$ is said to be
\emph{isolated} if its intrinsic geometry is `time-independent',
i.e., satisfies ${\cal L}_\ell q = 0$ and ${\cal L}_\ell dV = 0$.
Since these conditions are all local to $\IH$, the notion of an IH
is free of the global and teleological peculiarities of event
horizons.

Symmetries of an IH are diffeomorphisms of $\IH$ which preserve
its geometry. By the very definition of an IH, the null normal
$\ell$ is a symmetry. The question is whether there are others. A
complete classification of the symmetry groups is available
\cite{abl2}. If an IH admits 3 rotational symmetries
---so that $q$ is the metric of a round sphere--- it is said to
be of type I. If it admits an axial symmetry, it is said to be of
type II. Note that these symmetries refer to the IH itself; they
need not extend to space-time. Physically, type II IHs are the
most interesting ones.\\

\emph{Multipoles} \cite{aepv}: Because the notion of an IH is
quasi-local, Kerr event horizons constitute only a small sub-class
of type II IHs. More general IH geometries may be distorted. A
diffeomorphism invariant characterization of the geometry is
provided by a set of mass and angular momentum multipole moments
$M_n,J_n$. The physical dimension of these quantities depends on
$n$. But they are completely determined by simpler, dimensionless
`geometric multipoles' $I_n,L_n$ and the horizon area $a$. In the
Kerr family, $I_n,L_n$ are functions only of two parameters $a,
J$, while in general they are arbitrary, subject only to some mild
regularity conditions \cite{aepv}. Here, we will emphasize the
ideas and structures that are important to quantum theory.

For simplicity, let us suppose that there are no matter fields on
the horizon. Then, $I_n, L_n$ are the moments of the Weyl tensor
component $\Psi_2$ on $\Delta$ which is related to $V$ by
\be \label{dv} dV = \Psi_2 \,\,\epsilon\, ,  \ee
where $\epsilon$ is the area 2-form defined by $q$. The definition
of isolation implies that $\Psi_2$ is gauge invariant. To
construct the moments, one needs a notion of spherical harmonics.
It turns out that associated with every type II IH geometry $(q,
V)$, there is a \emph{canonical} type I geometry $(\nt{q},
\nt{V})$. One uses the spherical harmonics defined by the round
metric $\nt{q}$. To define $\nt{q}$, let us first introduce a
function $\zeta$ on $S$. Consider the orbits of the axial Killing
field $\phi^a$ of $q_{ab}$ on $S$ and label its orbits by any
fiducial coordinate $z$. Then, $\zeta$ is given by the partial
areas:
\be \label{zeta} \zeta(z) = -1 + 2\, \f{a(z)}{a}\, , \ee
where $a$ is the total area of $S$ (with respect to $q_{ab}$) and
$a(z)$ the partial area from the south pole up to the orbit of
$\phi^a$ labelled by $z$. If $\varphi$ is the angular coordinate
along orbits of $\phi^a$, the desired round metric $\nt{q}$ is now
given by:
\be\label{qo} \nt{q}_{ab} = R^2(\nt{f}^{-1} D_a\zeta\, D_b\zeta +
\nt{f} D_a\varphi\, D_b\varphi)\, , \ee
where $R$ is the area radius, $a = 4\pi R^2$, and $\nt{f} =
1-\zeta^2$. It can be put in the standard form by setting $\zeta =
\cos\theta$. \emph{It turns out that $q$ and $\nt{q}$ share the
same area element.} Finally, the geometric multipoles are defined
by:
\be \label{multi} I_n + i L_n = - \oint_S \Psi_2\, Y^0_n(\zeta)\,
\d^2S \ee
where $S$ is any 2-sphere cross-section of $\Delta$. (Since all
geometries under consideration are axi-symmetric, we need only the
$Y^m_n$ with $m=0$.) The $I_n,L_n$ so defined are subject to mild
algebraic restrictions. If $(q,V)$ and $(\bar{q}, \bar{V})$ are
related by a diffeomorphism, they have the same area and
multipoles.

Next, let us consider the converse. Suppose we are given an area
$a$ and a set of numbers $I_n,L_n$ (satisfying the mild algebraic
conditions). Then one can construct, uniquely up to
diffeomorphisms, a type II horizon geometry $(q,V)$, with horizon
area $a$ and geometric multipoles $I_n,L_n$. Start with a fiducial
type I geometry $(\nt{q}, \nt{V})$, and set:
\be \label{psi21} \Psi_2 (\zeta) = \f{1}{R^2}\,
\sum_{n=o}^{\infty}\, (I_n + i L_n)\, Y^0_n(\zeta), \ee
where $\zeta$ and the weighting functions $Y_n^0(\zeta)$ are
defined by $\nt{q}$. The desired physical metric $q$ has the same
form as (\ref{qo}) with $\nt{f}(\zeta)$ replaced by $f(\zeta)$,
constructed from the real part of $\Psi_2$:
\be \label{f} f(\zeta):= 4\,\left[R^2_\IH\, \int_{-1}^\zeta \d
\zeta_1 \int_{-1}^{\zeta_1} \d \zeta_2 \, {\rm Re}\Psi_2
\,(\zeta_2)\right]\,\, +\,2(\zeta+1)\, . \ee
Finally, using full $\Psi_2$ of (\ref{psi21}) and the fiducial
type I connection $\nt{V}$, one can construct the desired ${\rm
U(1)}$ connection $V$:
\be \label{v} V_a = \nt{V}_a -\f{1}{4} (f^\prime - \nt{f}^\prime)
D_a\varphi + \frac{i}{2}\omega_a \, ,\ee
where the prime denotes derivative with respect to $\zeta$ and
$\omega_a$ is the unique 1-form satisfying $d\omega = {\rm
Im}\,\Psi_2\, \nt{\epsilon}$ and $\nt{q}^{ab} \nt{D}_a \omega_b
=0$. All the information about multipoles is coded in $V-\nt{V}$.
Consequently, given a fixed area $a$ and a set of multipoles
$I_n,L_n$, the condition (\ref{dv}) satisfied by $V$ is equivalent
just to the equation
\be \label{hbc} d\nt{V} = - \,\f{2\pi}{a}\,\, \epsilon \ee
satisfied by any type I spin-connection $\nt{V}$. This fact will
be useful later. The horizon geometry $(q, V)$ so constructed is
of type II and has the desired area and multipoles. Furthermore,
as one might expect, the fiducial $(\nt{q}, \nt{V})$ we began with
is the canonical type I geometry associated with $(q, V)$.
Finally, a detailed examination of this construction shows that to
obtain $(q,V)$, full knowledge of the type I pair $(\nt{q},
\nt{V})$ is not necessary; all one needs is a foliation, $\nt{V}$,
horizon area $a$ and the set of fixed multipoles $I_n,L_n$.
Therefore, as we will see, the above constructions extend to quantum
geometry. \\

\emph{Hamiltonian framework}: Fix a 3-manifold $M$ with an
internal boundary $S$ which has the topology of a 2-sphere. $M$ is
to be thought of as a (partial) Cauchy surface in $(\man,
g_{ab})$, and the internal boundary $S$, as the intersection of
$M$ with an isolated horizon $\IH$ in $\man$. The physical meaning
of various fields is more transparent in the self-dual rather than
real connection variables. Therefore, it is convenient to begin
with pairs $(A_a^i,\, \Sigma_{ab}^i)$ of forms with values in the
complexification of the Lie algebra ${\rm su(2)}$ on a 3-manifold
$M$. $A_a^i$ can be thought of as the pull-back to $M$ of a
self-dual Lorentz connection in space-time, and  $\Sigma_{ab}^i$,
the Hodge-dual of an orthonormal triad $E^a_i$ of density weight 1
on $M$ (apart from a factor of $16\pi i\, G$. For conventions, see
\cite{alrev}.) $\epsilon_{ab}:= (16\pi i\, G)
\underline{\Sigma}_{ab}^i\,r_i$ is the area 2-form on $S$ induced
by $E^a_i$, where the underbar denotes pull-back to $S$ and $r^i$
is the internal radial vector. $V_a := \frac{1}{2}
\underbar{A}_a^i\, r_i$ on $S$ is the pull-back to $S$ of the
complex-valued ${\rm U(1)}$ connection of the horizon geometry.
Thus, the real part of $V_a$ is a $U(1)$ connection on the
spin-bundle over $S$ while the imaginary part is a 1-form
potential for the 2-form ${\rm Im}\Psi_2\, \epsilon$.

Our phase space ${\bf \Gamma}$ will consists of smooth pairs
$(A_a^i, \Sigma_{ab}^i)$ subject to certain boundary conditions.
At infinity, the fields fall off suitably to be asymptotically
flat (or, in the presence of a negative cosmological constant,
asymptotically anti-de Sitter). For our purposes, the important
boundary conditions are at $S$ and they ensure that we are
restricting ourselves to space-times with type II horizons having
fixed area and multipoles. We require that: i) the induced
geometry $(q, V)$ on $S$ is axi-symmetric with respect to
\emph{some} axial field $\phi^a$; ii) has a \emph{fixed} area
$a_o$; and iii) the $\Psi_2$ defined through (\ref{dv}) leads to a
\emph{fixed} set of multipoles ${I}^o_n,{L}^o_n$ via
(\ref{multi}).%
\footnote{In the type I case, all physical multipoles except $M_o$
are zero, $M_o$ is simply $\sqrt{a_o/16\pi}$, and $\Psi_2$ of
(\ref{dv}) is given by $-2\pi/a_o$. Therefore to single out the
relevant sector of general relativity, it suffices to fix just the
horizon area $a_o$. This is precisely what was done in \cite{abck}
although at that time the notion of multipoles was not available.}
One can show that, on this restricted phase space, the symplectic
structure of \cite{afk,abl2} reduces to:
\be \label{sym1} {\bf \Omega}(\delta_1, \delta_2) = -\int_M \,
{\rm Tr}\, (\delta_1 A \wedge \delta_2 \Sigma - \delta_2 A \wedge
\delta_1 \Sigma) \, +\, \f{1}{2\pi}\,\f{a_o}{4\pi G i } \oint_S \,
\delta_1 \nt{V} \wedge \delta_2 \nt{V}\,\, , \ee
where $\delta_1, \delta_2$ are any two tangent vectors to the phase
space ${\bf \Gamma}$;  ${\rm Tr}$ denotes the trace on internal
${\rm su(2)}$ indices; and $\nt{V}$ is the canonical type I spin
connection associated with $(q,V)$ via (\ref{v}).

Finally, the functional calculus on the space of connections is
well-developed only when the holonomies take values in compact Lie
groups. With self-dual connections, in the Lorentzian domain this
condition is not met. Therefore, as usual \cite{alrev}, we will
make a canonical transformation to real ${\rm SU(2)}$ connections,
essentially by replacing the $i$ in the relation between the
canonical and geometrical variables with a real, positive
parameter $\gamma$, called the Barbero-Immirzi parameter: $(A^i,\,
\S^i) \mapsto ({}^\g\!A^i,\, {}^\g\S^i )$. In terms of these
variables, the full symplectic structure is given by:
\be \label{sym2} {\bf \Omega}(\delta_1, \delta_2) = -  \int_M \,
{\rm  Tr}\, (\delta_1\, {}^\g\!A \wedge \delta_2 \,{}^\g \Sigma -
\delta_2\, {}^\g\!A \wedge \delta_1\, {}^\g \Sigma) \, +\,
\f{1}{2\pi}\, \f{a_o}{4\pi \g\, G} \oint_S \, \delta_1 \nt{V} \wedge
\delta_2 \nt{V}\, .  \ee
Note that the surface term is the symplectic structure of a
\emph{Chern-Simons theory} for an ${\rm U(1)}$ connection
$\nt{V}$, with level $k = a_o/4\pi\g \lp^2$. Because variations of
only $\nt{V}$ ---rather than ${}^\g V:= {}^\g\! \underbar{A}^i
r_i$--- appear, (\ref{hbc}) provides the most convenient way to
incorporate the key boundary condition that $dV$ is given by
(\ref{dv}), where $\Psi_2$ has the given set of
multipoles.\\

\emph{Quantum horizon geometry}: Each type II horizon geometry
defines a canonical type I geometry. Furthermore, the Hamiltonian
theory of the sector of general relativity admitting type II
isolated horizons \emph{with fixed area and multipoles} makes
direct reference only to the type I connection $\nt{V}$ and the
restriction to fixed multipole sector is now coded in condition
(\ref{hbc}). Therefore, the problem of quantization reduces
to that of type I theory and we can just take over all the
mathematical constructions from \cite{abk}. However, the physical
interpretation of states and operators has to be made in terms of
the physical type II geometries now under consideration.

Let us first summarize the mathematical structure from \cite{abk}.
To begin with, there is a kinematical Hilbert space $\H = \H_B
\otimes \H_S$ where $\H_B$ is built from suitable functions of
generalized connections in the bulk and $\H_S$ from suitable
functions of generalized surface connections on $S$. The bulk
Hilbert space $\H_B$ describes the `polymer excitations' of the
bulk geometry. Each excitation which punctures $S$ endows it with
a certain quantum of area. The surface Hilbert space $H_S$
consists of states of the level $k$, ${\rm U(1)}$ Chern-Simons
theory for the connection $\nt{V}$ on the punctured $S$. To ensure
that $S$ is indeed the desired horizon, only those states in $\H$
are selected which satisfy the operator analog of (\ref{hbc}),
called the quantum horizon boundary condition.%
\footnote{In addition, the states have quantum area and multipoles
close to the macroscopic classical values. The multipole operators
are defined below.}
This operator equation on permissible states allows the connection
and the triad to fluctuate but demands that they do so in tandem.
As emphasized in \cite{abk}, this operator equation in stringent
and admits a sufficient number of solutions only because of a
surprising agreement between an infinite set of eigenvalues of a
quantum geometry operator in the bulk agree with an infinite set
of eigenvalues of a Chern-Simons operator on $S$ which in turn is
possible precisely because of the isolated horizon boundary
conditions. The subspace $\Hk$ of $\H$ on which this condition is
met is then the Hilbert space of kinematic states describing
quantum geometry in the sector of general relativity now under
consideration.

To describe the quantum horizon geometry, the first step is to use
an operator analog of the partial area coordinate $\zeta$ which
played a key role in the classical theory. Introduce a fiducial
`foliation' of $S$ using \textit{some} axial field $\phi^a$ and
introduce a coordinate $z$ to label the leaves. Motivated by
(\ref{zeta}), we are led to define as operator $\hat{\zeta}$ on
$\Hk$:
\be \label{zetahat} \hat{\zeta}(z) = -1 + 2\,
\f{\hat{a}(z)}{\hat{a}_S}\, , \ee
where $\hat{a}_S$ is the area operator associated with $S$ and
$\hat{a}(z)$ is the area operator associated with the open portion
$S_z$ of $S$ bounded by the orbit of $\phi^a$ labelled by $z$ (and
containing, say, the `south pole'). To make the action of this
operator explicit, let us first note \cite{abk} that the Hilbert
space $\Hk$ can be decomposed as a direct sum,
\be \Hk = \oplus_{{\cal P},\, \vec{j}} \H^{{\cal P}, \vec{j}}, \ee
where ${\cal P}$ denotes a finite set of punctures and $\vec{j}$
is a set of half integers labelling the punctures. In each state
in $\H^{{\cal P},\, \vec{j}}$  the $I$th puncture is endowed with
a quantum of area of magnitude  $8\pi\g\, \sqrt{j_I(j_{I}+1)}
\lp^2$. Each of these subspaces is an eigenspace of the
$\hat{\zeta}$ operator, with eigenvalue:
\be \zeta^{{\cal P},\, \vec{j}}(z)  = -1 +2 \frac{\sum_{I'}\,
\sqrt{j_{I'} (j_{I'} +1)}}{\sum_I \sqrt{j_I (j_I +1)}}\, ,\ee
where the sum in the numerator ranges over all punctures on $S_z$
while the sum in the denominator ranges over all punctures on $S$.
(The presence of $\hat{a}_S$ ---rather than $a_o$--- in the
denominator ensures that eigenvalues of $\hat{\zeta}$ range from
$-1$ to $1$ as required in the definition of
$Y_n^0(\hat{\zeta})$.) In the classical theory, the knowledge of
$\zeta$ and multipoles ${I}^o_n, {L}^o_n$ suffices to determine
the horizon geometry $(q, V)$. The idea is to mimic that strategy.
However, care is needed because the eigenvalues of $\hat{\zeta}$
are discontinuous functions: they jump at each $z$ value where the
orbit of $\phi^a$ contains a puncture. This makes the quantum
geometry `rough'.

Fix a state in $\H^{{\cal P},\, \vec{j}}$. To make the nature of
the quantum geometry in this state explicit, let us introduce a
set of smooth functions $\zeta_\k (z)$ on $S$ which converge to
the eigenvalue $\zeta^{{\cal P},\, \vec{j}}(z)$ in the sup norm as
$k$ tends to infinity. Then, each $\zeta_\k$ defines via
(\ref{qo}) a round metric $\nt{q}_\k$. Using the fixed multipoles,
for each $k$, we can define a smooth function $\Psi_2^\k$:
\be \label{psi22} \Psi_2^{(k)}(\zeta_{(k)}) = \f{1}{R_o^2}\,
\sum_{n=o}^{\infty}\, ({I}^o_n + i {L}^o_n)\, Y^0_n(\zeta_{(k)}).
\ee
Using ${\rm Re}\Psi_2$ in (\ref{f}), we obtain a sequence of
functions $f^\k, \nt{f}^\k$ through (\ref{f}) and hence a sequence
of axi-symmetric metrics $q_\k$ and round metrics $\nt{q}_\k$ via
(\ref{qo}). As $k$ tends to infinity, $\Psi_2^\k,\,f^\k$ and
$\nt{f}^\k$ have well-defined limits $\Psi_2, f, \nt{f}$ which,
however, are discontinuous functions on $S$. However, $q_\k,
\nt{q}_\k$ do \emph{not} admit limits even in the distributional
sense because the metric coefficients are quadratic in
${d\zeta_\k}/{dz}$ and these functions tend to Dirac distributions
in the limit. This is not surprising because polymer quantum
geometry does not naturally admit metric operators. Nonetheless,
one can regard the family $q_\k$ as providing an intuitive
visualization of the quantum metric on the horizon in the
following sense. First, a type II metric is completely determined
by multipoles and the function $\zeta$, and in the above
construction multipoles are fixed and the $\zeta_\k$ tend to the
physical $\zeta$ uniformly. Second, every type II metric
determines the multipoles $I_n$ and for the family $q_\k$ these
are precisely the given ${I}^o_n$.

As one might expect from the type I analysis \cite{abk}, the
quantum connection operator can be defined more directly. Since
$\hat{\nt{V}}$ is a well-defined quantum connection on $\H_S$,
using (\ref{v}) we can define an operator $\hat{V}$ on $\Hk$:
\be \label{vhat} \hat{V} = \hat{\nt{V}} -\f{1}{4}  (f^\prime -
\nt{f}^\prime)(\hat{\zeta})\, D_a\varphi +
\frac{i}{2}\omega_a(\hat{\zeta})\, , \ee
where $f^\prime, \nt{f}^\prime$ and $\omega$ are all defined by the
limiting procedure described above. $\hat{V}$ is a well-defined
quantum connection: One can show that its holonomies along
arbitrary (analytic) edges on $S$, including those which may have
a puncture at their end points, are well-defined. $\hat{\nt{V}}$
is flat everywhere except at the punctures in the sense that the
holonomy around a closed loop not enclosing any puncture is
identity. This is not the case with $\hat{V}$. The distortion in
the quantum horizon geometry manifests itself through these
non-trivial holonomies. However, the `quantum Gauss Bonnet theorem'
\cite{abk} continues to hold: $\exp i\oint_S d\hat{V} = 1$.

Finally, we can define multipoles operators. Taking the limit $k
\rightarrow \infty$ of (\ref{psi22}) we obtain the $\Psi_2$
operator corresponding to the fiducial foliation of $S$:
\be \label{psi2hat} \hat{\Psi}_2 (z) = \f{1}{R_o^2}\,
\sum_{n=o}^{\infty}\, ({I}^o_n + i {L}^o_n)\,
Y^0_n(\hat{\zeta}(z)). \ee
(The numerical coefficient is left ${1}/{R_o^2}$ --- rather than
${4\pi}/\hat{a}_S$ --- to ensure the agreement with the definition
of $\hat{\Psi}_2$ used in quantum horizon boundary condition of
the type I analysis.) Quantum multipoles can be defined by
replacing $\zeta$ in (\ref{multi}) by $\hat\zeta$. However some
care involving a regularization in terms of $\zeta_\k$ is needed
to give precise meaning to the integral. One finds:
\be \label{multihat} \hat{I}_n + i \hat{L}_n := \lim_{k \to
\infty} -\oint_S \Psi_2^\k Y_n^0(\zeta_\k)\,\d^2S_\k
=\frac{\hat{a}_S}{a_o}\, ({I}^o_n + i{L}^o_n)\, ,\ee
where $\hat{a}_S$ is the total area operator associated with $S$.
Recall that in the analysis of type I horizons, the area was fixed
classically but could have small quantum fluctuations in quantum
theory. In the type II case the situation is similar with
multipoles. Multipoles are the moments in the spherical harmonic
decomposition with respect to $\hat\zeta$ and it is the quantum
fluctuations in $\hat{\zeta}$ that induce quantum fluctuations in
$\hat{I}_n, \hat{L}_n$. Since the former are dictated by
fluctuations in $\hat{a}_S$ so are the latter. Finally, in this
calculation the presence of $\hat{a}_S$ in the numerator is
dictated by the definition (\ref{zetahat}) of $\hat{\zeta}$ while
that of $a_o$ in the denominator comes from the definition
(\ref{psi2hat}) of $\hat{\Psi}_2$.

\emph{Entropy}: The calculation of entropy can be taken over from
the type I analysis in a straightforward fashion. (Indeed, most of
the above discussion of quantum operators encoding type II horizon
quantum geometry is inessential to the counting argument.) We
first impose quantum Einstein equations following the same
procedure as in \cite{abk}. Denote the resulting Hilbert space by
$\Hp$. To incorporate the fact that we are interested in the
horizon states of a black hole with fixed parameters, let us
construct a micro-canonical ensemble consisting of states in $\Hp$
for which the horizon area and multipoles lie in a small interval
around $a_o, {I}^o_n, {L}^o_n$ and count the Chern-Simons surface
states in this ensemble. Since eigenstates of $\hat{a}_S$ are also
eigenstates of $\hat{I}_n, \hat{L}_n$ and eigenvalues of
$\hat{I}_n, \hat{L}_n$ are completely determined by ${I}^o_n,
{L}^o_n$ and $a_S$, the counting is the same as in the type I case
\cite{dl,km}. Hence the entropy $S_{\rm hor}$ is again given by
\be \label{S} S_{\rm hor} =  \frac{a_{\rm hor}}{4\lp^2} -
\frac{1}{2}\, \ln(\frac{a_{\rm hor}}{\lp^2}) + {o}\,
\ln(\frac{a_{\rm hor}}{\lp^2})  \ee
provided $\gamma$ is chosen as in the type I analysis \cite{km}.

We will conclude with a comment on inclusion of matter fields. If
matter is minimally coupled to gravity, as in the type I case,
there are no matter surface terms in the symplectic structure,
whence there are no independent \emph{surface} degrees of freedom
associated with these matter fields. Furthermore, the
gravitational symplectic structure continues to be given by
(\ref{sym2}) whence the analysis summarized here undergoes only
inessential changes. In the non-minimally coupled case, the
gravitational symplectic structure does change but, by introducing
multipoles also for the matter fields, one can extend the analysis
of \cite{ac} and show that the classically expected entropy
expression \cite{acs} is recovered again for the same value of the
Barbero-Immirzi parameter.

Proofs, details of constructions and a discussion of merits and
limitations of this framework will appear elsewhere.

\section*{Acknowledgments}
We thank Jacek Wi\'{s}niewski, Alex Perez and Daniel Sudarsky for
stimulating discussions. This work was supported in part by the
National Science Foundation grants PHY-0090091, the Eberly
research funds of Penn State and the Alexander von Humboldt
Foundation of Germany.

\end{document}